\documentclass
[superscriptaddress,secnumarabic,amssymb,amsmath,nobibnotes,aps,prd,showkeys,showpacs,nofootinbib,onecolumn,notitlepage,12pt]{revtex4}%
\usepackage{setspace}
\usepackage{color}
\usepackage{amsmath}
\usepackage{amsfonts}
\usepackage{verbatim}
\usepackage{amssymb}
\usepackage{graphicx,bm}
\usepackage{graphicx}
\usepackage{bbm}
\usepackage{amsmath}
\usepackage{amssymb}
\usepackage{bbm}
\usepackage{amssymb}
\usepackage{graphicx,bm}
\usepackage{graphicx}
\usepackage{bbm}
\usepackage{epstopdf}
\usepackage[caption=false]{subfig}
\usepackage{makecell}%
\providecommand{\U}[1]{\protect\rule{.1in}{.1in}}
\newcommand{\be}{\begin{equation}}
\newcommand{\ee}{\end{equation}}

\newcommand{\mincir}{\raise
-3.truept\hbox{\rlap{\hbox{$\sim$}}\raise4.truept\hbox{$<$}\ }}
\newcommand{\magcir}{\raise
-3.truept\hbox{\rlap{\hbox{$\sim$}}\raise4.truept\hbox{$>$}\ }}

\ifx\pdfoutput\relax\let\pdfoutput=\undefined\fi
\newcount\msipdfoutput
\ifx\pdfoutput\undefined\else
\ifcase\pdfoutput\else
\ifx\paperwidth\undefined\else
\ifdim\paperheight=0pt\relax\else\pdfpageheight\paperheight\fi
\ifdim\paperwidth=0pt\relax\else\pdfpagewidth\paperwidth\fi
\fi\fi\fi
\begin{document}
\title{Dynamical analysis of $f\left(  Q\right)  $-cosmology}
\author{Andronikos Paliathanasis}
\email{anpaliat@phys.uoa.gr}
\affiliation{Institute of Systems Science, Durban University of Technology, Durban 4000,
South Africa}
\affiliation{Departamento de Matem\'{a}ticas, Universidad Cat\'{o}lica del Norte, Avda.
Angamos 0610, Casilla 1280 Antofagasta, Chile}

\begin{abstract}
We study the evolution of the physical variables in $f\left(  Q\right)
$-gravity for two families of symmetric and flat connections in a spatially
flat Friedmann--Lema\^{\i}tre--Robertson--Walker geometry where the equation
of motion for the nonmetricity scalar is not trivially identity. From the
analysis of dynamics we found that the de Sitter universe is always an
attractor while the cosmological models admit scaling solutions which can
describe the early acceleration phase of our universe or the matter and the
radiation epochs.

\end{abstract}
\maketitle
\date{\today}

\section{Introduction}

\label{sec1}

Symmetric teleparallel theory \cite{Nester:1998mp} is a gravitational theory
equivalent to General Relativity where the Levi-Civita connection is replaced
by a torsion-free and flat connection and the fundamental scalar for the
gravitational Lagrangian is the non-metricity scalar $Q$. Indeed, General
Relativity, symmetric teleparallel theory, and teleparallel equivalence of
General Relativity form a family of theories known as trinity of gravity
\cite{tr1}, in which the\ gravitational Lagrangian is a linear function of the
Ricci scalar $R$, the non-metricity scalar $Q$ and the torsion scalar $T$
respectively. The equivalency between these three theories is lost when
non-linear components of the above scalars are introduced in the gravitational Lagrangian.

The $f\left(  X\right)  -$theories of gravity where $X$ is a geometric scalar,
have been proposed by cosmologists \cite{mod1} in order to explain the
observational phenomena \cite{od1,od2,od3}, like the late-time acceleration
phase of the universe attributed to the dark matter. In $f\left(  X\right)
-$theories of gravity, the dark energy has a geometric origin \cite{sdc1}.
Indeed, the new degrees of freedom which are provided in the field equations
by the nonlinear components of the $f$ function define an effective
geometrodynamic fluid which can drive the dynamics in order to explain the
acceleration of the universe. When the three scalar of the trinity of gravity
are used on the $f\left(  X\right)  $-theories we end with the $f\left(
R\right)  ~$\cite{Buda}, the teleparallel $f\left(  T\right)  ~$%
\cite{Ferraro}, and the symmetric teleparallel $f\left(  Q\right)  $ \cite{f6}
gravitational theories. $f\left(  R\right)  $ and $f\left(  T\right)  $
gravitational models have been widely studied in the literature for
cosmological studies \cite{rt1,rt2,rt3,rt4,rt5,rt6}, and for the analysis of
astrophysical objects \cite{rt7,rt8,rt9,rt10}; for more details we refer the
reader to the review articles \cite{rt11,rt12}.

In this work we focus on the cosmological application of $f\left(  Q\right)
$-gravity in a isotropic and homogeneous spatially flat
Friedmann--Lema\^{\i}tre--Robertson--Walker background. In symmetric
teleparallel theory of gravity it is possible to separate the inertial effects
from gravity because through an appropriate coordinate transformation we can
define the coincident gauge where all the components of the flat connect
vanish. Only the recent years $f\left(  Q\right)  $-gravity has drawn the
attention of cosmologists. In \cite{ff1,ff2,ff3,ff4,ff4a,ff4b} $f\left(
Q\right)  $-gravity has been considered as a geometric dark energy model;
furthermore in \cite{ff5} the authors show that $f\left(  Q\right)  $-gravity
could challenge the $\Lambda$CDM model. An inflationary scenario in $f\left(
Q\right)  $-gravity is presented in \cite{ff6}. Anisotropic cosmological
models were investigated in \cite{an1,an2} while inhomogeneous spacetimes were
studied in \cite{Heis1,ww4,ww5}. For other recent studies of $f\left(
Q\right)  $-gravity we refer the reader to
\cite{nn1,nn2,nn3,nn4,nn5,nn6,nn7,nn8} and references therein.

We perform a detailed analysis for the phase-space of the field equations in
$f\left(  Q\right)  $-gravity for different families of flat connections.
These different families of connections provide the same field equations in
symmetric teleparallel equivalence of general relativity, however for
nonlinear functions $f\left(  Q\right)  $ there are introduced dynamical
degrees of freedom from the connection. In order to understand the effects of
these dynamical terms in the cosmological parameters we determine the
asymptotic solutions. This approach has been widely applied in gravitational
theories with many important results
\cite{df1,df2,df3,df4,df5,df6,df7,df8,df9,df10}. From this analysis we can
construct constraints for the free parameters of the given theory and for the
initial value problem \cite{rt2}. The structure of the paper is as follows.

In Section \ref{sec2} we present the gravitational model of our consideration
which is that of symmetric teleparallel $f\left(  Q\right)  $-gravity. For the
background space we assume isotropic and homogeneous spatially flat
Friedmann--Lema\^{\i}tre--Robertson--Walker (FLRW) spacetime. We present the
three different families of symmetric and flat connections and the
corresponding field equations. For two families of connections, an arbitrary
scalar is introduced in the field equations. As a result the corresponding
field equations admit a different dynamical evolution. We focus our study on
these two connections. The main results of this work are presented in Section
\ref{sec3} where we perform a detailed analysis of the phase-space for the
field equations. We apply a set of dimensionless variables and we determine
the admitted asymptotic solutions by the cosmological field equations.
Finally, in Section \ref{sec4} we summarize our results and we draw our conclusions.

\section{$f\left(  Q\right)  $-gravity}

\label{sec2}

In symmetric teleparallel $f\left(  Q\right)  -$gravity the gravitational
Action Integral is \cite{Zhao}
\begin{equation}
S=\frac{1}{2}\int d^{4}x\sqrt{-g}f(Q)+\int d^{4}x\sqrt{-g}\mathcal{L}%
_{M}+\lambda_{\kappa}^{\;\lambda\mu\nu}R_{\;\lambda\mu\nu}^{\kappa}%
+\tau_{\lambda}^{\;\mu\nu}\mathrm{T}_{\;\mu\nu}^{\lambda}, \label{action}%
\end{equation}
where $g=\mathrm{det}(g_{\mu\nu})$, $g_{\mu\nu}$ is the metric tensor,
quantities $R_{\;\lambda\mu\nu}^{\kappa}$,~$\mathrm{T}_{\;\mu\nu}^{\lambda}$
are the Riemann and torsion tensors respectively, $\lambda_{\kappa}%
^{\;\lambda\mu\nu}$, $\tau_{\lambda}^{\;\mu\nu}$ are two Lagrange multipliers
which impose that the connection $\Gamma_{\mu\nu}^{\kappa}$ is flat and
symmetric, that is, $R_{\;\lambda\mu\nu}^{\kappa}=0$ and $\mathrm{T}_{\;\mu
\nu}^{\lambda}=0$.

$f\left(  Q\right)  $ is an arbitrary function of the non-metricity scalar $Q$
for the connection $\Gamma_{\mu\nu}^{\kappa}$ in which
\begin{equation}
Q=Q_{\lambda\mu\nu}P^{\lambda\mu\nu},
\end{equation}
with
\begin{equation}
Q_{\lambda\mu\nu}=\frac{\partial g_{\mu\nu}}{\partial x^{\lambda}}%
-\Gamma_{\;\lambda\mu}^{\sigma}g_{\sigma\nu}-\Gamma_{\;\lambda\nu}^{\sigma
}g_{\mu\sigma},
\end{equation}
and $P_{\;\mu\nu}^{\lambda}$ to be the non-metricity conjugate tensor defined
as%
\begin{equation}
P_{\;\mu\nu}^{\lambda}=-\frac{1}{4}Q_{\;\mu\nu}^{\lambda}+\frac{1}{2}%
Q_{(\mu\phantom{\lambda}\nu)}^{\phantom{(\mu}\lambda\phantom{\nu)}}+\frac
{1}{4}\left(  Q^{\lambda}-\bar{Q}^{\lambda}\right)  g_{\mu\nu}-\frac{1}%
{4}\delta_{\;(\mu}^{\lambda}Q_{\nu)},
\end{equation}
where $Q_{\lambda}=Q_{\lambda~~~\mu}^{~~~\mu}$, $\bar{Q}_{\lambda
}=Q_{~~\lambda\mu}^{\mu}$.

In the case of vacuum, variation with respect to the metric tensor of the
gravitational Action Integral provides the field equations
\begin{equation}
\frac{2}{\sqrt{-g}}\nabla_{\lambda}\left(  \sqrt{-g}f^{\prime}(Q)P_{\;\mu\nu
}^{\lambda}\right)  -\frac{1}{2}f(Q)g_{\mu\nu}+f^{\prime}(Q)\left(  P_{\mu
\rho\sigma}Q_{\nu}^{\;\rho\sigma}-2Q_{\rho\sigma\mu}P_{\phantom{\rho\sigma}\nu
}^{\rho\sigma}\right)  =0, \label{feq1a}%
\end{equation}
where now prime denotes a derivative with respect to the non-metricity scalar,
i.e. $f^{\prime}(Q)=\frac{df\left(  Q\right)  }{dQ}$. Equivalently, the field
equations (\ref{feq1a}) can be written with the use of the Einstein-tensor
$G_{\mu\nu}$ as follow%
\begin{equation}
f^{\prime}(Q)G_{\mu\nu}+\frac{1}{2}g_{\mu\nu}\left(  f^{\prime}%
(Q)Q-f(Q)\right)  +2f^{\prime\prime}(Q)\left(  \nabla_{\lambda}Q\right)
P_{\;\mu\nu}^{\lambda}=0, \label{feq1}%
\end{equation}
where now $G_{\mu\nu}=\tilde{R}_{\mu\nu}-\frac{1}{2}g_{\mu\nu}\tilde{R},$ with
$\tilde{R}_{\mu\nu}$ and $\tilde{R}$ are the Riemannian Ricci tensor and
scalar respectively which are constructed by the Levi-Civita connection.
Indeed, when $f\left(  Q\right)  $ is a linear function, the field equations
(\ref{feq1}) reduce to that of Genereal Relativity.

Over and above , from the variation of the Action Integral with respect to the
connection we derive the equation of motion for the non-metricity scalar%
\begin{equation}
\nabla_{\mu}\nabla_{\nu}\left(  \sqrt{-g}f^{\prime}%
(Q)P_{\phantom{\mu\nu}\sigma}^{\mu\nu}\right)  =0. \label{feq2}%
\end{equation}

\subsection{FLRW background}

In this study for the background geometry we consider the spatially flat FLRW
geometry with scale factor $a\left(  t\right)  $ and line element
\begin{equation}
ds^{2}=-dt^{2}+a(t)^{2}\left[  dr^{2}+r^{2}\left(  d\theta^{2}+\sin^{2}\theta
d\phi^{2}\right)  \right]  ,
\end{equation}
where now the Hubble function is defined as $H=\frac{\dot{a}}{a}$, in which a
dot means total derivative with respect to the time-parameter $t$, i.e.
$\dot{a}=\frac{da}{dt}$.

The connections which are symmetric, flat and inherit the six isometries of
the FLRW as collineations have been determined before in \cite{Hohmann,Heis2}%
.\ Indeed, there exist three family of connections with common components%
\begin{align*}
\Gamma_{\theta\theta}^{r}  &  =-r~,~\Gamma_{\phi\phi}^{r}=-r\sin^{2}\theta~,\\
\Gamma_{r\theta}^{\theta}  &  =\Gamma_{\theta r}^{\theta}=\Gamma_{r\phi}%
^{\phi}=\Gamma_{\phi r}^{\phi}=\frac{1}{r}\\
\Gamma_{\phi\phi}^{\theta}  &  =-\sin\theta\cos\theta~,~\Gamma_{\theta\phi
}^{\phi}=\Gamma_{\phi\theta}^{\phi}=\cot\theta.
\end{align*}

The first connection, namely $\Gamma_{1}$ has the additional non-zero
component%
\[
\Gamma_{\;tt}^{t}=\gamma(t).
\]

Furthermore, the second-connection, $\Gamma_{2},$ has the additional non-zero
components%
\[
\Gamma_{\;tt}^{t}=\frac{\dot{\gamma}(t)}{\gamma(t)}+\gamma(t),\quad
\Gamma_{\;tr}^{r}=\Gamma_{\;rt}^{r}=\Gamma_{\;t\theta}^{\theta}=\Gamma
_{\;\theta t}^{\theta}=\Gamma_{\;t\phi}^{\phi}=\Gamma_{\;\phi t}^{\phi}%
=\gamma(t).
\]

Finally, the third connection $\Gamma_{3}$ has the extra non-zero components%
\[
\Gamma_{\;tt}^{t}=-\frac{\dot{\gamma}(t)}{\gamma(t)},\quad\Gamma_{\;rr}%
^{t}=\gamma(t),\quad\Gamma_{\;\theta\theta}^{t}=\gamma(t)r^{2},\quad
\Gamma_{\;\phi\phi}^{t}=\gamma(t)r^{2}\sin^{2}\theta.
\]
where function $\gamma$ is an arbitrary functions of the time-parameter $t$.

We remark that connections $\Gamma_{1}$, $\Gamma_{2}$ and $\Gamma_{3}$ are the
same only for zero function $\gamma\left(  t\right)  $. Moreover, for each
connection the field equations (\ref{feq1}) are different.

Indeed, for the first connection $\Gamma_{1}$ we derive the non-metricity
scalar $Q=-6H^{2}$ and the field equations are
\begin{align}
3H^{2}f^{\prime}(Q)+\frac{1}{2}\left(  f(Q)-Qf^{\prime}(Q)\right)   &
=0~,\label{cc.01}\\
2\frac{d}{dt}\left(  f^{\prime}(Q)H\right)   &  =0,\label{cc.02}%
\end{align}
where equation (\ref{feq2}) is satisfied as identity. From (\ref{cc.02}) it
follows that $\sqrt{Q}f\,^{\prime}\left(  Q\right)  =const$. The latter is an
algebraic equation from where it follows that $Q$ is always a constant
parameter, that is, $H$ is also a constant parameter, i.e. $H=H_{0}$. Thus, in
the case of vacuum for connection $\Gamma_{1}$, the only possible solution is
the de Sitter universe, $H_{0}\neq0$, or the Minkowski spacetime, $H_{0}=0$.
For the power-law function $f\left(  Q\right)  =Q^{\alpha}$, equation
(\ref{cc.01}) gives that that the unique solution is the Minkowski scacetime
where $H=0$.

For the second connection $\Gamma_{2}$ we calculate the non-metricity scalar
$Q=-6H^{2}+9\gamma H+3\dot{\gamma}$, while equations (\ref{feq1}) and
(\ref{feq2}) read
\begin{align}
3H^{2}f^{\prime}(Q)+\frac{1}{2}\left(  f(Q)-Qf^{\prime}(Q)\right)
+\frac{3\gamma\dot{Q}f^{\prime\prime}(Q)}{2} &  =0,\label{cc.03}\\
2\frac{d}{dt}\left(  f^{\prime}(Q)H\right)  -3\gamma\dot{Q}f^{\prime\prime}(Q)
&  =0,\label{cc.04}\\
\dot{Q}^{2}f^{\prime\prime\prime}(Q)+\left[  \ddot{Q}+3H\dot{Q}\right]
f^{\prime\prime}(Q) &  =0.\label{cc.05}%
\end{align}
From the latter equations it is clear that function $\gamma\left(  t\right)  $
plays an important role on the dynamical evolution of the field equations.

On the other-hand, for the third connection $\Gamma_{3}$ the non-metricity
scalar is $Q=-6H^{2}+\frac{3\gamma}{a^{2}}H+\frac{3\dot{\gamma}}{a^{2}},$ and
the field equations are%
\begin{align}
3H^{2}f^{\prime}(Q)+\frac{1}{2}\left(  f(Q)-Qf^{\prime}(Q)\right)
-\frac{3\gamma\dot{Q}f^{\prime\prime}(Q)}{2a^{2}}  &  =0,\label{cc.06}\\
2\frac{d}{dt}\left(  f^{\prime}(Q)H\right)  +\frac{\gamma\dot{Q}%
f^{\prime\prime}(Q)}{a^{2}}  &  =0,\label{cc.07}\\
\dot{Q}^{2}f^{\prime\prime\prime}(Q)+\left[  \ddot{Q}+\dot{Q}\left(
H+\frac{2\dot{\gamma}}{\gamma}\right)  \right]  f^{\prime\prime}(Q)  &  =0.
\label{cc.08}%
\end{align}
where again function $\gamma\left(  t\right)  $ effects the dynamics of the
physical parameters.

\section{Dynamical analysis}

\label{sec3}

We proceed our analysis with the analysis of phase-space for the cosmological
field equations for the $f\left(  Q\right)  $-gravity. The dynamical analysis
for the first connection with a matter source performed in \cite{dyn1} where
it was found that the universe admits as asymptotic solutions the de Sitter
universe, and the matter dominated eras. A similar analysis appeared later in
\cite{dyn2}. Indeed, when there is not any external matter source for the
first connection the only possible asymptotic solution is that of the de
Sitter universe. While this work was in preparation the authors in \cite{dyn3}
performed an analysis for the phase-space of the field equations for the
second connection $\Gamma_{2}$ with matter source. For this case new
stationary points exist where the scalar $\gamma\left(  t\right)  $
contributes in the cosmological fluid. \newline

Indeed, the existence of self-similar solutions was the subject of study in
\cite{self} and it was found that for connections $\Gamma_{2}$ and $\Gamma
_{3}$ self-similar solutions exist when a power-law component $Q^{\alpha}$
dominates in the $f\left(  Q\right)  $ function. Thus for simplicity of our
analysis we consider the power-law function $f\left(  Q\right)  $ and we
investigate the dynamics of the field equations in order to reconstruct the
evolution of the physical parameters.

We consider the power-law $f\left(  Q\right)  =Q^{\alpha}$ function with
$\alpha\neq0,1$ and we define new dimensionless variables in the context of
the $H$-normalization approach. We rewrite the field equations as a system of
first-order differential equations and we determine the stationary points.
Each stationary point describes an asymptotic solution for the background
space. Furthermore we investigate the stability properties of the stationary
points in order to reconstruct the cosmological history provided by the theory.

\subsection{Connection $\Gamma_{2}$}

In order to study the dynamical evolution for the field equations of the
second connection for the power-law theory $f\left(  Q\right)  =Q^{\alpha},$
we introduce the new variables%
\[
x=\frac{\alpha-1}{6\alpha}\frac{Q}{H^{2}}~,~y=\frac{\left(  1-\alpha\right)
}{2}\frac{\gamma}{H^{2}}\frac{\dot{Q}}{Q}~,~z=\frac{\gamma}{H}~,~\tau=\ln a.
\]

Hence, the field equations (\ref{cc.03}), (\ref{cc.04}) and (\ref{cc.05}) are
expressed as follow%
\begin{align}
\frac{dx}{d\tau}  &  =\frac{2}{\alpha-1}\frac{xy}{z}\left(  1-2\alpha+3\left(
\alpha-1\right)  z\right)  ,\label{cc.09}\\
\frac{dy}{d\tau}  &  =\frac{2}{\alpha-1}\frac{y}{z}\left(  \alpha x+\left(
\alpha-1\right)  \left(  y-1\right)  \left(  3z-1\right)  \right)
,\label{cc.10}\\
\frac{dz}{d\tau}  &  =\frac{2\alpha}{\alpha-1}x+\left(  y-1\right)  \left(
3z-2\right)  , \label{cc.11}%
\end{align}
with algebraic constraint%
\begin{equation}
1-x-y=0. \label{cc.12}%
\end{equation}

Moreover, the equation of state parameter $w_{eff}=-1-\frac{2}{3}\frac{\dot
{H}}{H^{2}}$ for the effective cosmological fluid reads%
\begin{equation}
w_{eff}\left(  x,y,z\right)  =-1+y\left(  2-\frac{4}{3z}\right)  .
\label{cc.13}%
\end{equation}

It is important to mention the variables $x,~y$ and $z$ are constraint and
they can take values in all the space of real numbers. Indeed, they can take
values at the infinity. Furthermore, with the use of the algebraic equation
(\ref{cc.12}) we can reduce the dimension by one the dimension of the
dynamical system. Thus, in the two-dimensional plane $\left(  x,z\right)  $
the dynamical system reads%
\begin{align}
\frac{dx}{d\tau}  &  =\frac{2}{\alpha-1}\frac{x\left(  1-x\right)  }{z}\left(
1-2\alpha+3\left(  \alpha-1\right)  z\right)  ,\label{cc.14}\\
\frac{dz}{d\tau}  &  =x\left(  \frac{2\alpha}{\alpha-1}-\left(  3z-2\right)
\right)  . \label{cc.15}%
\end{align}
We note that $z$ is constraint as $z>0$ or $z\,<0$.

The stationary points $A=\left(  x\left(  A\right)  ,z\left(  A\right)
\right)  $ of the latter dynamical system at the finite regime are
\[
A_{1}=\left(  1,\frac{2\left(  2\alpha-1\right)  }{3\left(  \alpha-1\right)
}\right)  ~,~A_{2}=\left(  0,z_{2}\right)  .
\]

For point $A_{1}$ we calculate $w_{eff}\left(  A_{1}\right)  =-1$, which means
that the asymptotic solution is that of the de Sitter universe. On the
other-hand $A_{2}$ describes a family of points with $w_{eff}\left(
A_{2}\right)  =1-\frac{4}{3z_{2}}$, which means that the asymptotic solutions
at points $A_{2}$ describe accelerated universes for $0<z_{2}<1$, while for
$z_{2}=\frac{2}{3}$ the de Sitter universe is recovered.

\begin{figure}[ptb]
\centering\includegraphics[width=1\textwidth]{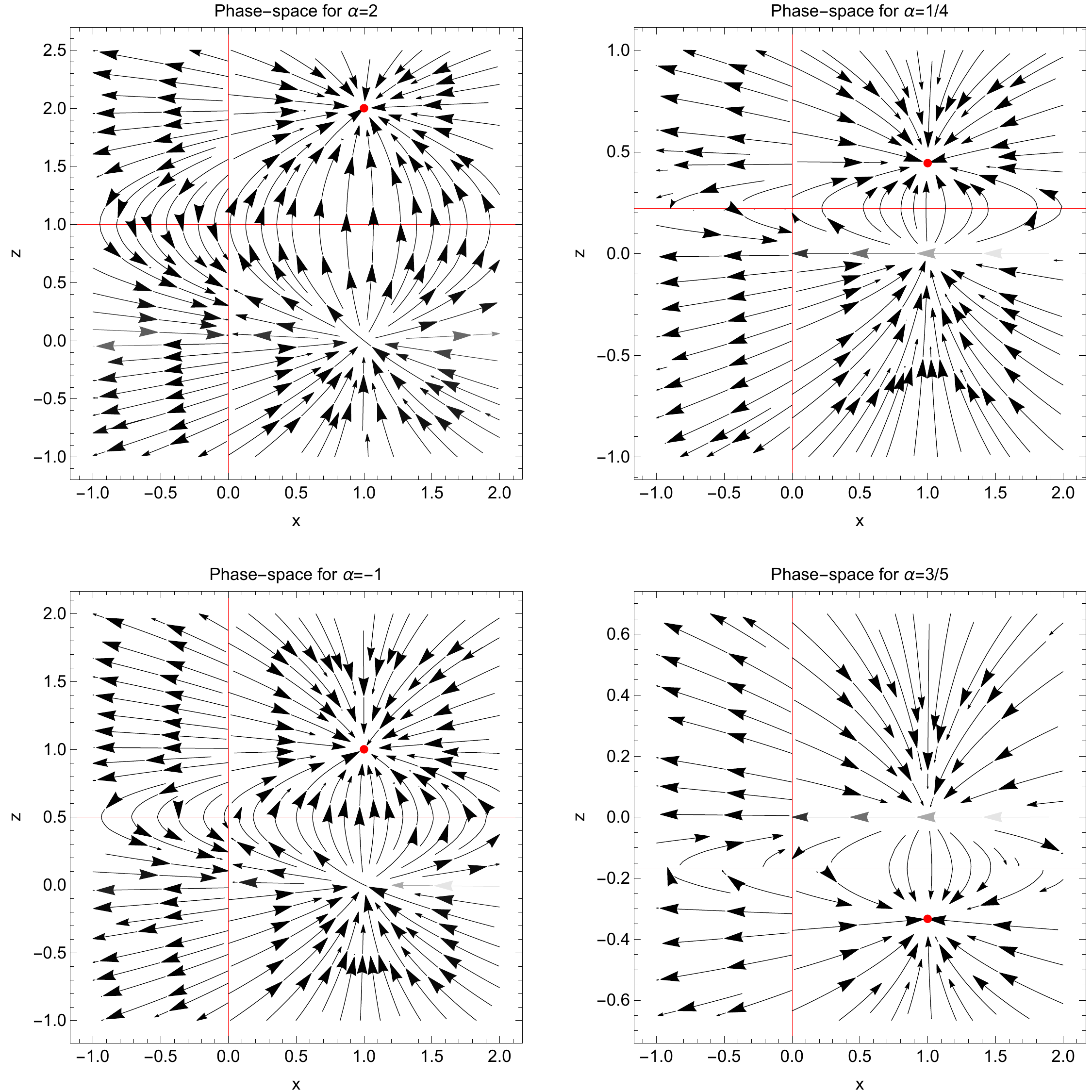}\caption{Phase-space
portrait for the dynamical system (\ref{cc.14}), (\ref{cc.15}) for different
values of parameter $\alpha$. It is clear that the family of stationary points
admit a stable submanifold. Moreover, point $\left(  x,z\right)  =\left(
1,0\right)  $ is a transition point where parameter $z$ changes sign. }%
\label{fig1}%
\end{figure}

As far as the stability of the stationary points is concerned for the
linearized system (\ref{cc.14}), (\ref{cc.15}) near the stationary point
$A_{1}$ we derive the two eigenvalues $e_{1}\left(  A_{1}\right)  =-3$ and
$e_{2}\left(  A_{2}\right)  =-3$ from where we infer that the de Sitter
universe is always a future attractor. On the other hand, for the family of
points $A_{2}$ we determine the eigenvalues $e_{1}\left(  A_{2}\right)  =0$
and $e_{2}\left(  A_{2}\right)  =\frac{2\left(  1-2\alpha-3z\left(
1+\alpha\right)  \right)  }{\left(  a-1\right)  z}$. When $e_{2}\left(
A_{2}\right)  <0,$ that is, for $\left\{  \alpha<\frac{1}{2},0<z_{2}%
<\frac{2\alpha-1}{3\left(  \alpha-1\right)  }\right\}  ,~\left\{  \frac{1}%
{2}<\alpha<1\text{, }\frac{2\alpha-1}{3\left(  \alpha-1\right)  }%
<z_{2}<0\right\}  $and $\left\{  \alpha>1,0<z_{2}<\frac{2\alpha-1}{3\left(
\alpha-1\right)  }\right\}  $, there may exist a stable submanifold. The
latter can be calculated with the application of the center manifold theorem.
However, such a discussion is a mathematical tool and does not contribute in
the physical discussion of the theory. Instead in Fig. \ref{fig1} we present
the phase-space portrait for the dynamical system (\ref{cc.14}), (\ref{cc.15})
in the finite regime. For the diagrams it is clear that stationary points
$A_{2}$ admit always a stable submanifold, however in general the family of
points are saddle points.

From \ref{fig1}, point $\bar{A}=\left(  1,0\right)  $ seems to be a stationary
points for the dynamical system (\ref{cc.14}), (\ref{cc.15}), however that is
not true, since equation (\ref{cc.15}) reads $\frac{dz}{d\tau}|_{\bar
{A}=\left(  1,0\right)  }=\frac{2\alpha}{\alpha-1}+2$. However, $\bar{A}$ is a
transition point where parameter $z$ change sing. 

\subsubsection{Analysis at Infinity}

In order to investigate the existence of stationary points at the infinity we
define the Poincare variables
\[
x=\frac{X}{\sqrt{1-X^{2}-Z^{2}}}~,~z=\frac{Z}{\sqrt{1-X^{2}-Z^{2}}}%
\text{,~}d\tau=\sqrt{1-X^{2}-Z^{2}}d\sigma\text{.}%
\]

Therefore, the system (\ref{cc.14}), (\ref{cc.15}) becomes
\begin{equation}
\frac{dX}{d\sigma}=F_{1}\left(  X,Z;\alpha\right)  ~;~\frac{dZ}{d\sigma}%
=F_{2}\left(  X,Z;\alpha\right)  \text{. }%
\end{equation}
Infinity is reached when $1-X^{2}-Z^{2}=0$. Hence the physical accepted
stationary points at infinity are
\[
A_{1}^{\infty}=\left(  0,1\right)  \text{ and }A_{2}^{\infty}=\left(
0,-1\right)  .
\]

\begin{figure}[ptb]
\centering\includegraphics[width=1\textwidth]{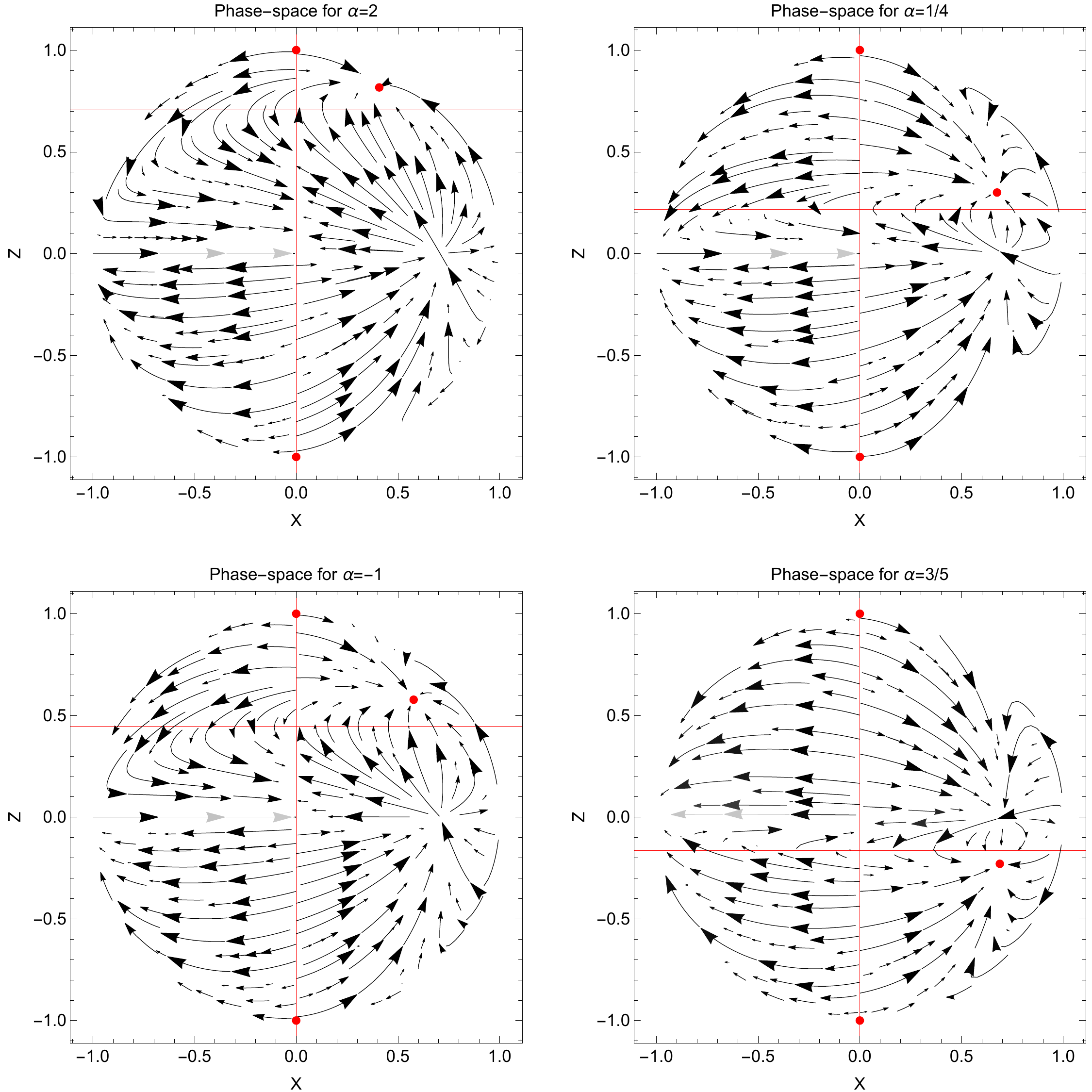}\caption{Phase-space
portrait for the dynamical system (\ref{cc.14}), (\ref{cc.15}) for different
values of parameter $\alpha$ at the Poincare variables. We observe that the de
Sitter point $A_{1}$ is the unique future attractor. }%
\label{fig2}%
\end{figure}

Points $A_{1}^{\infty}$ and $A_{2}^{\infty}$ describe scaling solutions with
$w_{eff}\left(  A_{1,2}^{\infty}\right)  =1$, that is, they describe universes
dominated by a stiff fluid. The two eigenvalues of the linearized system close
to the stationary points are determined to be zero. In Fig. \ref{fig2} we
present the phase-space portrait for the dynamical system in the Poincare
variables from where it is clear that the two stationary points at infinity
regime describe unstable solutions.

\begin{figure}[ptb]
\centering\includegraphics[width=1\textwidth]{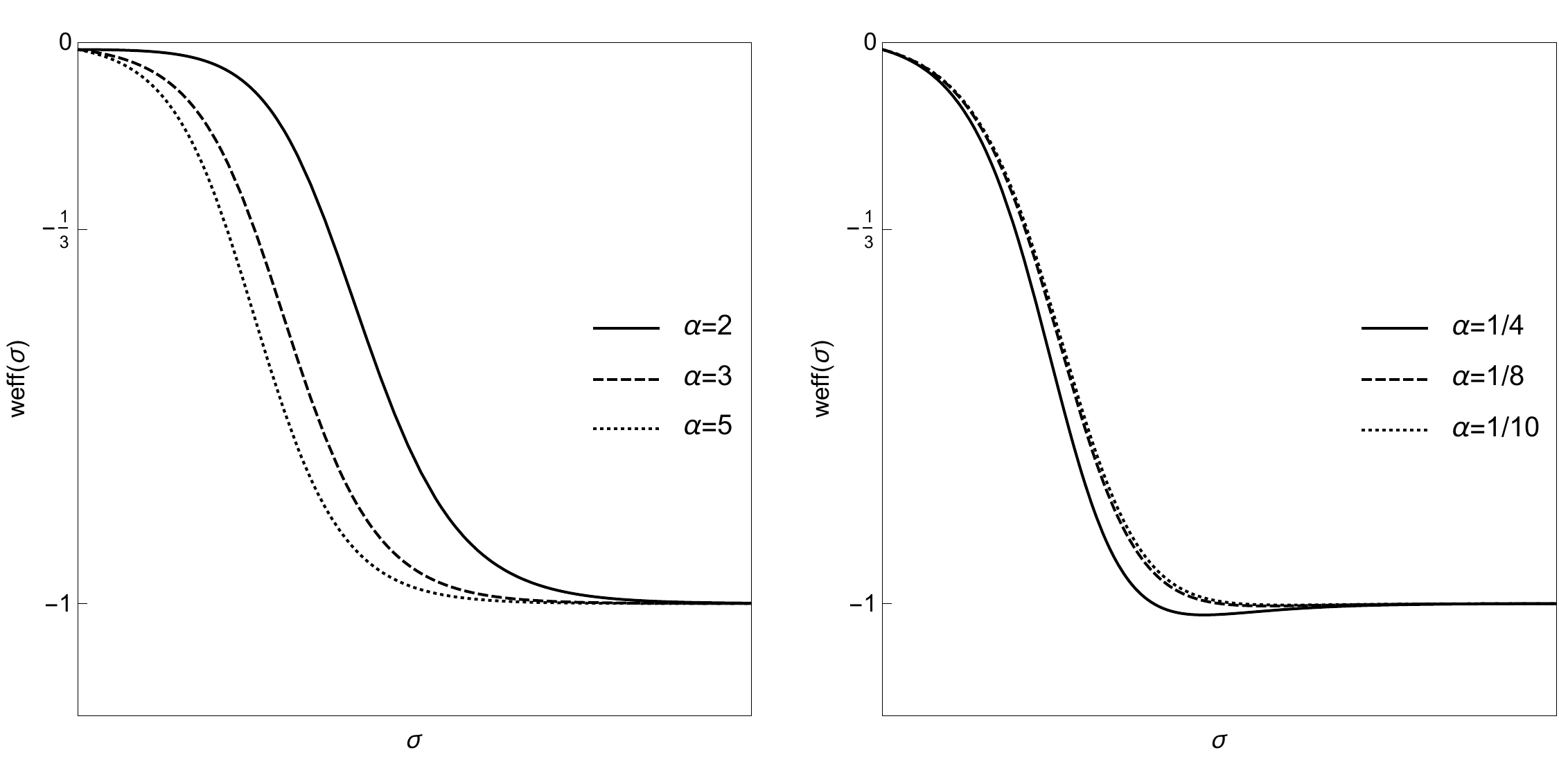}\caption{Qualitative
evolution for the effective equation of state parameter $w_{eff}\left(
\sigma\right)  ~$of connection $\Gamma_{2}$ for different values of the free
parameter $\alpha$ and for initial condition such that the matter dominated
era to be described by the effective fluid.}%
\label{fig3}%
\end{figure}

\begin{figure}[ptb]
\centering\includegraphics[width=0.5\textwidth]{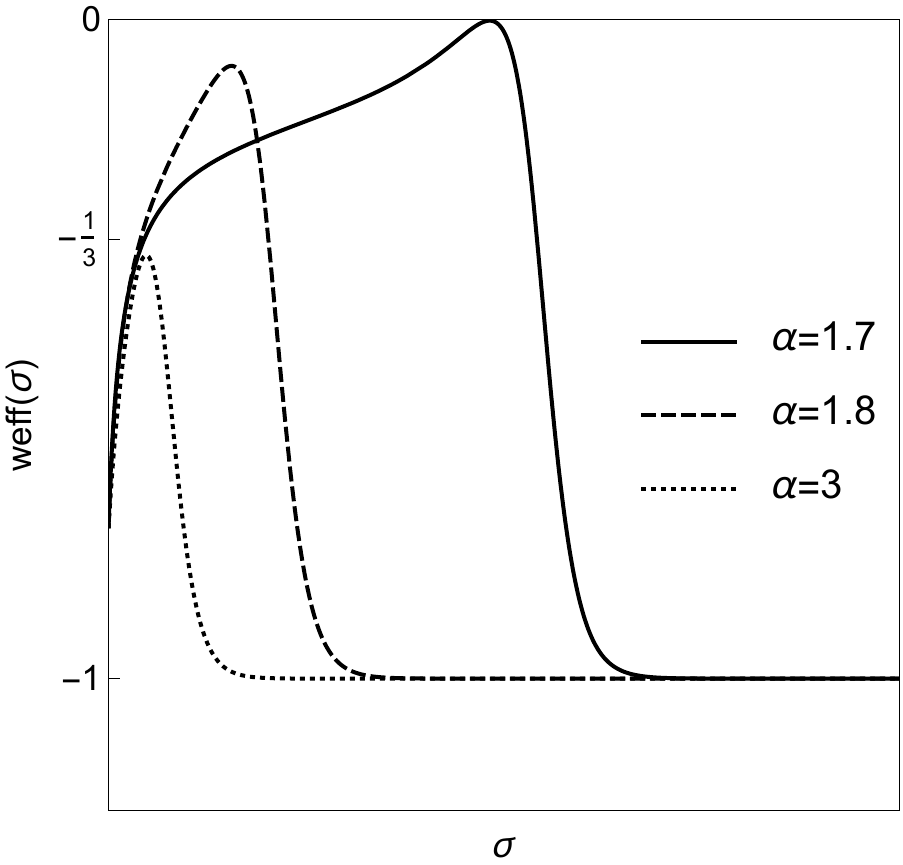}\caption{Qualitative
evolution for the effective equation of state parameter $w_{eff}\left(
\sigma\right)  $ of connection $\Gamma_{2}~$for different values of the free
parameter $\alpha$ and for initial condition such that to describe exit from
inflation.}%
\label{fig4}%
\end{figure}

In Fig. \ref{fig3} we present the qualitative evolution for the effective
equation of state parameter $w_{eff}\left(  \sigma\right)  $ for different
values of the free parameter $\alpha$ and for initial condition such that the
matter dominated era to be described by the effective fluid. We observe that
the power-law model $f\left(  Q\right)  =Q^{\alpha}$ for the second connection
can be used as unified model for the dark sector of the universe. On the other
hand, in Fig. \ref{fig4} we present the qualitative evolution of
$w_{eff}\left(  \sigma\right)  $ for different values of the free parameters
with initial conditions which describe an unstable accelerated universe. From
the qualitative evolution of $w_{eff}\left(  \sigma\right)  $ there exist
solution trajectories which describe universes with phases acceleration,
matter era and future acceleration. Consequently, this cosmological model can
be used as a toy model to unify the inflation with the late-time acceleration
and the matter era. 

As we discussed for the variables in the finite regime points with $Z=0$ and
$X=1$, are transition points and not stationary points.

\subsubsection{}

\subsection{Connection $\Gamma_{3}$}

For the third connection and the field equations (\ref{cc.06}), (\ref{cc.07})
and (\ref{cc.08}) we apply the same procedure. In this case we define the
dimensionless variables%
\[
\hat{x}=\frac{\alpha-1}{6\alpha}\frac{Q}{H^{2}}~,~\hat{y}=\frac{\left(
1-\alpha\right)  }{2}\frac{a^{2}\gamma}{H^{2}}\frac{\dot{Q}}{Q}~,~\hat
{z}=\frac{a^{2}\gamma}{H}~,~\tau=\ln a.
\]

In the new variables the field equations are%
\begin{align}
\frac{d\hat{x}}{d\tau}  &  =\frac{2}{\alpha-1}\frac{\hat{x}\hat{y}}{\hat{z}%
}\left(  1-2\alpha+\left(  \alpha-1\right)  \hat{z}\right)  ,\label{cc.20}\\
\frac{d\hat{y}}{d\tau}  &  =-\frac{2}{\alpha-1}\frac{\hat{y}}{\hat{z}}\left(
\alpha\hat{x}+\left(  \alpha-1\right)  \left(  1+\hat{y}\right)  \left(
1+\hat{z}\right)  \right)  ,\label{cc.21}\\
\frac{d\hat{z}}{d\tau}  &  =2\left(  1+\frac{\alpha}{\alpha-1}\hat{x}\right)
-3\hat{z}-\hat{y}\left(  2+\hat{z}\right)  , \label{cc.23}%
\end{align}
with algebraic equation%
\begin{equation}
1-\hat{x}+\hat{y}=0, \label{cc.24}%
\end{equation}
and $w_{eff}=-1-\frac{2}{3}\frac{\hat{y}}{\hat{z}}\left(  2+\hat{z}\right)  ,$
where $\hat{z}>0$ or $\hat{z}<0$.

With the use of the latter constraint equation we end with the two-dimensional
dynamical system%
\begin{align}
\frac{d\hat{x}}{d\tau}  &  =\frac{2}{\alpha-1}\frac{\hat{x}\left(  \hat
{x}-1\right)  }{\hat{z}}\left(  1-2\alpha+\left(  \alpha-1\right)  \hat
{z}\right)  ,\label{cc.25}\\
\frac{d\hat{z}}{d\tau}  &  =2\left(  1+\frac{\alpha}{\alpha-1}\hat{x}\right)
-3\hat{z}+\left(  1-\hat{x}\right)  \left(  2+\hat{z}\right)  . \label{cc.26}%
\end{align}

The stationary points $B=\left(  \hat{x}\left(  B\right)  ,\hat{z}\left(
B\right)  \right)  ~$for the dynamical system (\ref{cc.25}), (\ref{cc.26}) are%
\[
B_{1}=\left(  1,\frac{2\left(  2\alpha-1\right)  }{3\left(  \alpha-1\right)
}\right)  ~,~B_{2}=\left(  0,2\right)  \text{ and }B_{3}=\left(
\frac{2\left(  3-4\alpha\right)  }{1+2\alpha},\frac{1-2\alpha}{\alpha
-1}\right)  .
\]

For the stationary point $B_{1}$ we calculate $w_{eff}\left(  B_{1}\right)
=-1$ and the eigenvalues for the linearized system $e_{1}\left(  B_{1}\right)
=-3$ and $e_{2}\left(  B_{2}\right)  =-5$. Thus, the asymptotic solution at
point $B_{1}$ describes a de Sitter universe and point $B_{1}$ is a future
attractor. For point $B_{2}$ it follows $w_{eff}\left(  B_{2}\right)
=\frac{1}{3}$, $e_{1}\left(  B_{2}\right)  =-2$ and $e_{1}\left(
B_{2}\right)  =\frac{4\alpha-3}{\alpha-1}$; thus $B_{2}$ describes a universe
dominated by radiation and the point is an attractor for $\frac{3}{4}%
<\alpha<1$, otherwise is a saddle point. Finally, for the point $B_{3}$ we
calculate $w_{eff}\left(  B_{3}\right)  =\frac{7-6\alpha}{3+6\alpha}$,
$e_{\pm}\left(  B_{3}\right)  =1-\frac{5}{1+2\alpha}\pm\frac{2\sqrt
{\alpha\left(  41\alpha-14\right)  -11}}{1+2\alpha}$. Therefore point $B_{3}$
describe an accelerated universe for~$\alpha>2$, or $\alpha<-\frac{1}{2}$ and
the point is an attractor for $-\frac{1}{2}<\alpha<\frac{3}{4}$.

\begin{figure}[ptb]
\centering\includegraphics[width=1\textwidth]{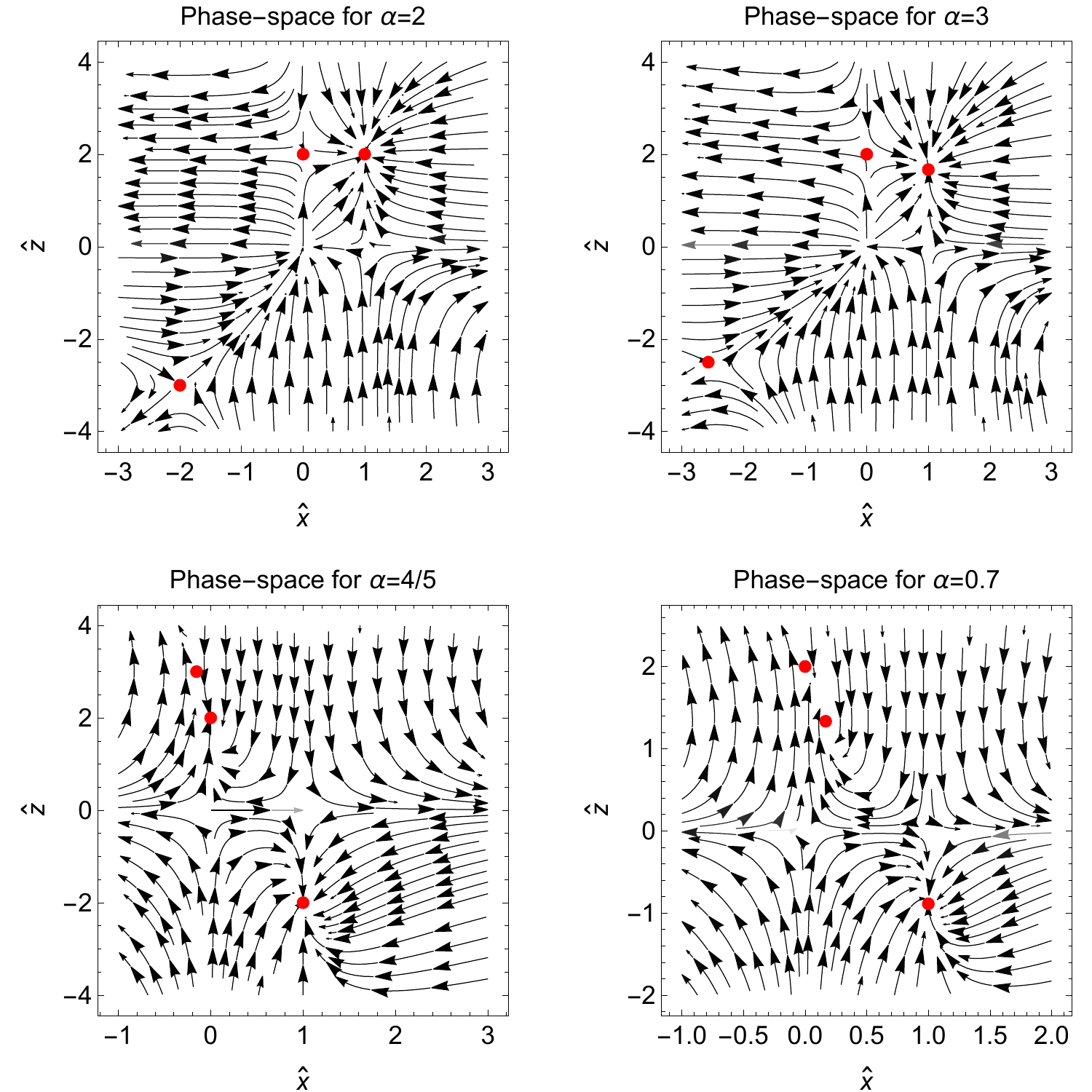}\caption{Phase-space
portrait for the dynamical system (\ref{cc.25}), (\ref{cc.26}) for different
values of parameter $\alpha$. The de Sitter point $B_{1}$ is always an
attractor.}%
\label{fig5}%
\end{figure}

In Fig. \ref{fig5} we present the phase-space portrait for the two-dimensional
system (\ref{cc.25}), (\ref{cc.26}) for various values of the free parameter
$\alpha$. 

It is important to mention that similarly with the dynamical system for the
connection $\Gamma_{2}$, point $\bar{B}=\left(  1,0\right)  $ is a transition
point and not a stationary point. 

\subsubsection{Analysis at Infinity}

We define the new Poincare variables%
\[
\hat{x}=\frac{\hat{X}}{\sqrt{1-\hat{X}^{2}-\hat{Z}^{2}}}~,~\hat{z}=\frac
{Z}{\sqrt{1-\hat{X}^{2}-\hat{Z}^{2}}}\text{,~}d\tau=\sqrt{1-\hat{X}^{2}%
-\hat{Z}^{2}}d\sigma\text{,}%
\]
thus the field equations (\ref{cc.25}), (\ref{cc.26}) are written in the form%
\begin{equation}
\frac{d\hat{X}}{d\sigma}=G_{1}\left(  \hat{X},\hat{Z};\alpha\right)
~;~\frac{d\hat{Z}}{d\sigma}=G_{2}\left(  \hat{X},\hat{Z};\alpha\right)
\text{. }%
\end{equation}

\begin{figure}[ptb]
\centering\includegraphics[width=1\textwidth]{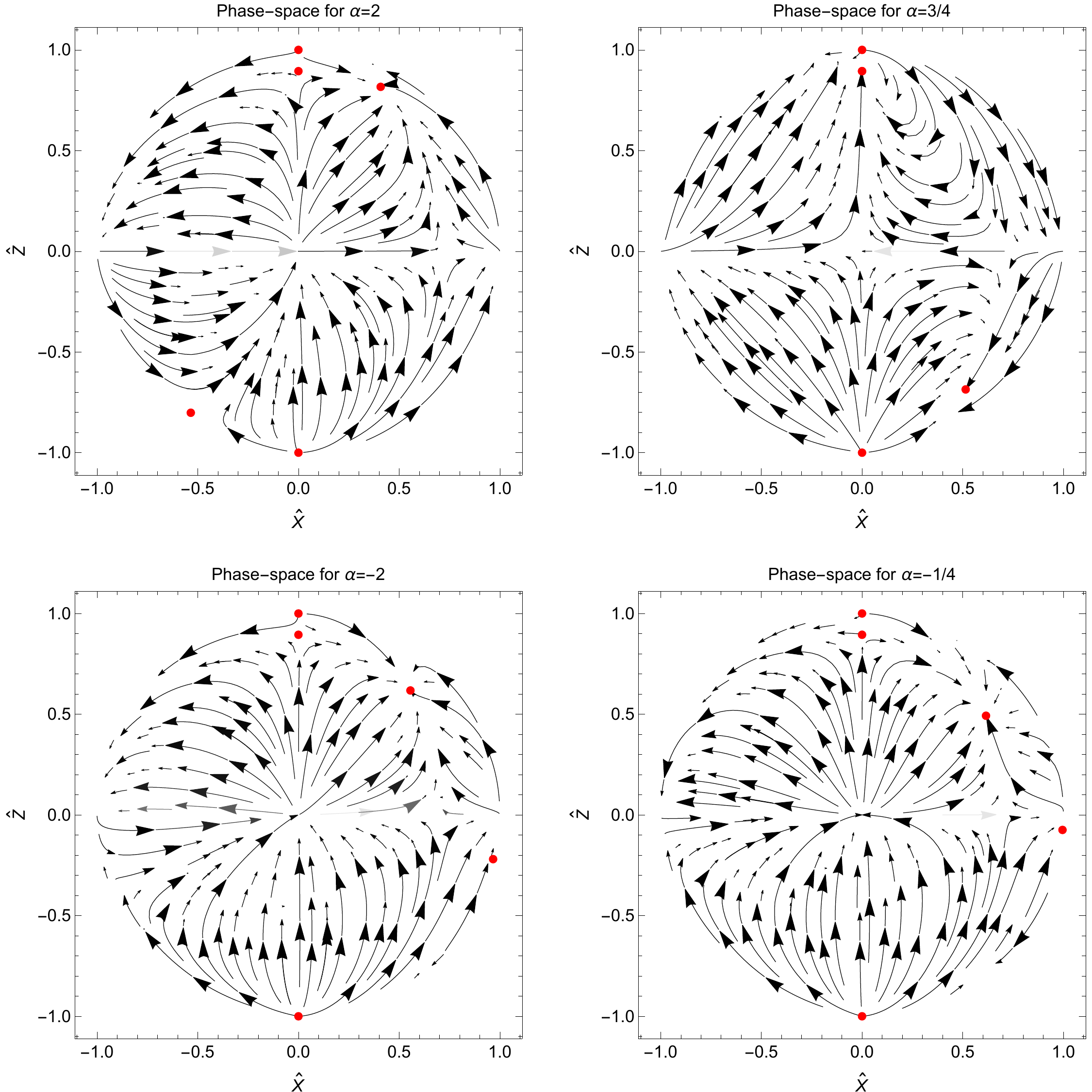}\caption{Phase-space
portrait for the dynamical system (\ref{cc.14}), (\ref{cc.15}) for different
values of parameter $\alpha$ at the Poincare variables. We observe that the
stationary points at the infinity regime are sources.}%
\label{fig6}%
\end{figure}

The stationary points of the latter dynamical system at the infinity, i.e. on
the surface $1-\hat{X}^{2}-\hat{Z}^{2}=0$, are
\[
B_{1}^{\infty}=\left(  0,1\right)  \text{ and }B_{2}^{\infty}=\left(
0,-1\right)  .
\]
The two points describe scaling solutions with $w_{eff}\left(  B_{1}^{\infty
}\right)  =w_{eff}\left(  B_{2}^{\infty}\right)  =-\frac{1}{3}.$ The
eigenvalues of the linearized system around the stationary points are zero.
From the\ phase-space portraits of Fig. \ref{fig6} we observe that the
stationary points $B_{1}^{\infty}$, $B_{2}^{\infty}$ are sources.

\begin{figure}[ptb]
\centering\includegraphics[width=1\textwidth]{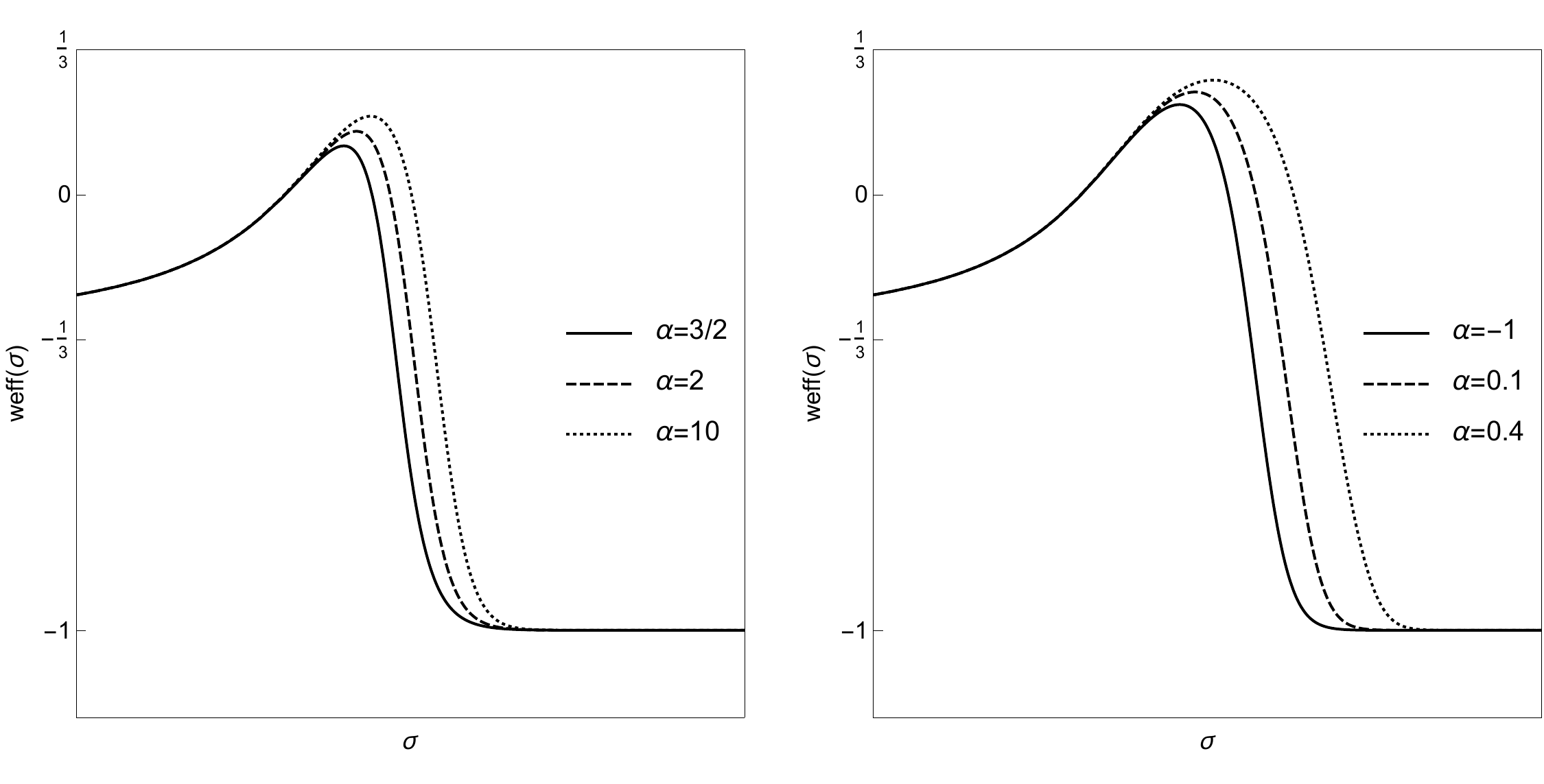}\caption{Qualitative
evolution for the effective equation of state parameter $w_{eff}\left(
\sigma\right)  $ of connection $\Gamma_{3}~$for different values of the free
parameter $\alpha$ and for initial condition such that to describe exit from
inflation.}%
\label{fig7}%
\end{figure}

In Fig. \ref{fig7} we plot the qualitative evolution of the equation of state
parameter $w_{eff}\left(  \sigma\right)  $ for initial conditions near
infinity. We remark that the $w_{eff}$ reaches the saddle point $B_{3}$ which
describes a radiation-like spacetime and then the de Sitter point $B_{1}$ is
the future attractor. Indeed, the evolution of the physical parameters for the
third connection can describe a radiation epoch as we exist from inflation and
and then to effective fluid has the behaviour of the cosmological constant.

The results of this Section for the dynamical analysis of the cosmological
field equations for the two connections $\Gamma_{2}$ and $\Gamma_{3}$ are
summarized in Table \ref{tab1}. \ %

%TCIMACRO{\TeXButton{B}{\begin{table}[tbp] \centering}}%
%BeginExpansion
\begin{table}[tbp] \centering
%EndExpansion
\caption{Stationary points and physical properties of the asymptotic solutions.}%
\begin{tabular}
[c]{cccccc}%
\multicolumn{6}{c}{\textbf{Connection }$\mathbf{\Gamma}_{2}$}\\\hline\hline
\textbf{Finite regime} & \textbf{Point} & $\left(  \mathbf{x,z}\right)  $ &
$\mathbf{w}_{eff}$ & \textbf{Acceleration?} & \textbf{Attractor?}\\\hline
& $A_{1}$ & $\left(  1,\frac{2\left(  2\alpha-1\right)  }{3\left(
\alpha-1\right)  }\right)  $ & $-1$ & \thinspace Yes & Yes\\
& $A_{2}$ & $\left(  0,z_{2}\right)  $ & $1-\frac{4}{3z_{2}}$ & $0<z_{2}<1$ &
No\\
&  &  &  &  & \\
\textbf{Infinity regime} & \textbf{Point} & $\left(  \mathbf{X,Z}\right)  $ &
$\mathbf{w}_{eff}$ & \textbf{Acceleration?} & \textbf{Attractor?}\\
& $A_{1}^{\infty}$ & $\left(  0,1\right)  $ & $1$ & No & No\\
& $A_{2}^{\infty}$ & $\left(  0,-1\right)  $ & $1$ & No & No\\\hline\hline
&  &  &  &  & \\
\multicolumn{6}{c}{\textbf{Connection }$\mathbf{\Gamma}_{3}$}\\\hline\hline
\textbf{Finite regime} & \textbf{Point} & $\left(  \mathbf{\hat{x},\hat{z}%
}\right)  $ & $\mathbf{w}_{eff}$ & \textbf{Acceleration?} &
\textbf{Attractor?}\\\hline
& $B_{1}$ & $\left(  1,\frac{2\left(  2\alpha-1\right)  }{3\left(
\alpha-1\right)  }\right)  $ & $-1$ & Yes & Yes\\
& $B_{2}$ & $\left(  0,2\right)  $ & $\frac{1}{3}$ & No & $\frac{3}{4}%
<\alpha<1$\\
& $B_{3}$ & $\left(  \frac{2\left(  3-4\alpha\right)  }{1+2\alpha}%
,\frac{1-2\alpha}{\alpha-1}\right)  $ & $\frac{7-6\alpha}{3+6\alpha}$ &
$\alpha>2$ or $\alpha<-\frac{1}{2}$ & $-\frac{1}{2}<\alpha<\frac{3}{4}$\\
&  &  &  &  & \\
\textbf{Infinity regime} & \textbf{Point} & $\left(  \mathbf{\hat{X},\hat{Z}%
}\right)  $ & $\mathbf{w}_{eff}$ & \textbf{Acceleration?} &
\textbf{Attractor?}\\
& $B_{1}^{\infty}$ & $\left(  0,1\right)  $ & $-\frac{1}{3}$ & No & No\\
& $B_{2}^{\infty}$ & $\left(  0,-1\right)  $ & $-\frac{1}{3}$ & No &
No\\\hline\hline
\end{tabular}
\label{tab1}%
%TCIMACRO{\TeXButton{E}{\end{table}}}%
%BeginExpansion
\end{table}%
%EndExpansion

\section{Conclusions}

\label{sec4}

In this study we investigated the asymptotic behaviour for the cosmological
parameters in symmetric teleparallel $f\left(  Q\right)  $-gravity for two
torsion-free and flat connections which provide non-trivial dynamical degrees
of the freedom in the field equations. We focused in the simplest case of
power-law $f\left(  Q\right)  =Q^{\alpha}$ theory and that there is not any
matter external matter source. While for the usual applied connection in the
literature such consideration leads to the Minkowski spacetime that is not
true for the two connections of the FLRW geometry that we consider in this work.

We defined new variables in the $H$-normalization approach and we wrote the
field equations in the equivalent form of two-dimensional dynamical system of
first-order differential equations. We determined the stationary points at the
finity and infinity regimes. The analysis at the infinity performed with the
introduction of Poincare variables. For both connections we found that the de
Sitter universe is always a future attractor for the theory. However, new
scaling solutions were determined at the finite and infinity regimes.

For the connection $\Gamma_{2}$ at the finite regime there exist a family
unstable scaling solutions which can describe acceleration, while at infinity
there exist stiff fluid solutions. On the other hand, for connection
$\Gamma_{2}$ at the finite regime there exists a stationary point which
describes an unstable radiation solution and a scaling solution which can
describe acceleration or matter era. At infinity the stationary points
describe solutions dominated by curvature-like fluid source.

In order to understand the evolution of the general solution we plotted
phase-space portraits to constrain the initial conditions from the evolution
of the trajectories in the two-dimensional space of the dynamical system. We
solved numerically the trajectories and from the qualitative evolution of the
effective equation of state parameter we found that the theory can explain the
evolution of the universe as we exit from the inflationary epoch and for
various sets of initial conditions it can be used as a toy model for the
description of the matter era and of the present acceleration.

In future work we plan to investigate further the physical properties of these
solutions of $f\left(  Q\right)  $-gravity for these two connections.

\begin{acknowledgments}
This work was financially supported in part by the National Research
Foundation of South Africa (Grant Numbers 131604). The author thanks the
support of Vicerrector\'{\i}a de Investigaci\'{o}n y Desarrollo
Tecnol\'{o}gico (Vridt) at Universidad Cat\'{o}lica del Norte through
N\'{u}cleo de Investigaci\'{o}n Geometr\'{\i}a Diferencial y Aplicaciones,
Resoluci\'{o}n Vridt No - 098/2022.
\end{acknowledgments}


\begin{thebibliography}{99}                                                                                               %


\bibitem {Nester:1998mp}M. Hohmann, Phys. Rev. D 104, 124077 (2021)

\bibitem {tr1}J. Beltr\'{a}n-Jim\'{e}nez, L. Heisenberg and T.S. Koivisto,
Universe 5, 173 (2019)

\bibitem {mod1}S. Nojiri, S.D. Odintsov and V.K. Oikonomou, Phys. Rept. 692, 1 (2017)

\bibitem {od1}A.G. Riess et al., Astron. J. 116, 1009 (1998)

\bibitem {od2}M. Tegmark et al., Astrophys. J. 606, 702 (2004)

\bibitem {od3}E. Komatsu et al., Astrophys. J. Suppl. Ser. 180, 330 (2009)

\bibitem {sdc1}S. Capozziello, C.A. Mantica and L.G. Molinari, EPL 137, 19001 (2022)

\bibitem {Buda}H.A. Buchdahl, Mon. Not. Roy. Astron. Soc. 150, 1 (1970)

\bibitem {Ferraro}R. Ferraro and F. Fiorini, Phys. Rev. D 75, 084031 (2007)

\bibitem {f6}J. B. Jimenez, L. Heisenberg and T. Koivisto, Phys. Rev. D 98,
044048 (2018)

\bibitem {rt1}S. M. Carrol, V. Duvvuri, M. Trodden and M. S. Turner, Phys.
Rev. D., 70, 043528 (2004);

\bibitem {rt2}L.~Amendola, D.~Polarski and S.~Tsujikawa,
Phys.\ Rev.\ Lett.\ 98, 131302 (2007)

\bibitem {rt3}A. A. Starobinsky, Phys. Lett. B 91, 99 (1980)

\bibitem {rt4}T. Clifton and J.D. Barrow, Class. Quant. Grav. 23, 2951 (2006)

\bibitem {rt5}R.C. Nunes, A. Bonilla, S. Pan and E.N. Saridakis, EPJC 77, 230 (2017)

\bibitem {rt6}S. Nesseris, S. Basilakos, E.N. Saridakis and L.
Perivolaropoulos, Phys. Rev. D 88, 103010 (2013)

\bibitem {rt7}A. de la Cruz-Dombriz, A. Dobado and A. L. Maroto, Phys. Rev. D
80, 124011 (2009)

\bibitem {rt8}E. Santos, Astroph. Sp. Sci. 341, 411 (2012)

\bibitem {rt9}M. E. Rodrigues, M. J. S. Houndjo, J. Tossa, D. Momeni and R.
Myrzakulov, JCAP 11, 024 (2013)

\bibitem {rt10}J. Aftergood and A. DeBenedictis, Phys. Rev.\ D 90, 124006 (2014)

\bibitem {rt11}T. P. Sotiriou and V. Faraoni, Rev. Mod. Phys. 82 451, (2010)

\bibitem {rt12}S. Bahamonde, K. F. Dialektopoulos,
C.\ Escamilla-Rivera,\ G.\ Farrugia, V. Gakis, M. Hohmann, J. L. Said, J.
Mifsud and E. Di Valentino, Rep. Prog. Phys. 86, 026901 (2023)

\bibitem {ff1}L. Atayde and N. Frusciante, Phys. Rev. D 104, 064052 (2021)

\bibitem {ff2}R. Solanki, A. De and P.K. Sahoo, Phys. Dark Universe 36, 100996 (2022)

\bibitem {ff3}A. Lymperis, JCAP 11, 018 (2022)

\bibitem {ff4}S.H. Shekh, Phys. Dark Univ. 33, 100850 (2021)

\bibitem {ff4a}S.A. Narawade and B. Mishra, Phantom Cosmological Model with
Observational Constraints in f(Q) Gravity, to appear in Annalen de Physik,
(2023) DOI: 10.1002/andp.202200626

\bibitem {ff4b}S.A. Narawade, Laxmipriya Pati, B. Mishra, S.K. Tripathy,
Physics of Dark Universe 36, 101020 (2022)

\bibitem {ff5}F.K. Anagnastopoulos, S.\ Basilakos and E.N. Saridakis, Phys.
Lett. B 822, 136634 (2021)

\bibitem {ff6}S. Capozziello and M. Shokri, Phys. Dark Univ. 37, 101113 (2022)

\bibitem {ww1}A.\ De, S. Mandal, J.T. Beh,\ T.-H. Loo and P.K. Sahoo, EPJC 82,
72 (2022)

\bibitem {ww2}W. Wang, H. Chen and T. Katsuragawa, Phys.\ Rev. D 105, 024060 (2022)

\bibitem {Heis1}F. D' Ambrosio, S. D. B. Fell, L. Heisenberg and S. Kuhn,
Phys. Rev. D 105, 024042 (2022)

\bibitem {ww4}R.-H. Lin and X.-H. Zhai, Phys. Rev. D 103, 124001 (2021)

\bibitem {ww5}S. Mandal, G. Mustafa, Z. Hassan and P.K.Sahoo, Phys. Dark
Universe 35, 100934 (2022)

\bibitem {qq1}K. Hu, T. Katsuragawa and T. Qiu, ADM formulation and
Hamiltonian analysis of f(Q) gravity, [arXiv:2204.12826] (2022)

\bibitem {qq2}N. Dimakis, A. Paliathanasis and T. Christodoulakis, Class.
Quantum\ Grav. 38, 225003 (2021)

\bibitem {an1}A. De, S. Mandal, J.T. Beh, T.-H. Loo and P.K.\ Sahoo, Eur.
Phys. J. C 82, 72 (2022)

\bibitem {an2}F. Esposito, S. Carloni and S. Vignolo, Class. Quantum Grav. 39,
235014 (2022)

\bibitem {nn1}N. Dimakis, A. Paliathanasis and T. Christodoulakis, Class.
Quantum\ Grav. 38, 225003 (2021)

\bibitem {nn2}M. Koussour, S. Dahmani, M. Bennai and T. Quali, Eur. Phys. J.
Plus 138, 179 (2023)

\bibitem {nn3}W. Khyllep, J. Dutta, E.N. Saridakis and K. Yesmakhanova, Phys.
Rev. D 107, 044022 (2023)

\bibitem {nn4}M. Hohmann, Phys.\ Rev. D 104, 124077 (2021)

\bibitem {nn5}K. Hu, T. Katsuragawa and T. Qiu, Phys. Rev. D 106, 044025 (2022)

\bibitem {nn6}G. Subramaniam, A. De, T.-H. Loo and Y.K. Goh, How different
connections in flat FLRW geometry impact energy conditions in f(Q) theory?,
(2023) [arXiv:2304.02300]

\bibitem {nn7}M Koussour, S.K.J. Pacif, M. Bennai and P.K. Sahoo, Fortschritte
der Physik 71, 2200172 (2023)

\bibitem {nn8}M. Koussour and M. Bennai, Chinese Journal of Physics 79, 339 (2022)

\bibitem {df1}A.A. Coley, Phys.\ Rev. D 62, 023517 (2000)

\bibitem {df2}R. Lazkoz and G. Leon, Phys. Lett. B 638, 303 (2006)

\bibitem {df3}J. De-Santiago, J.L. Cervantes-Cota and D. Wands, Phys. Rev. D
87, 023502 (2013)

\bibitem {df4}A. Cid, G. Leon and Y. Leyva, JCAP 02, 027 (2016)

\bibitem {df5}S. Chatzidakis, A. Giacomini, P.G.L. Leach, G. Leon, A.
Paliathanasis and S. Pan, JHEAp 36, 141 (2022)

\bibitem {df6}H.R. Kausar, I. Noureen and M.U. Shahzad, Eur. Phys. J. Plus
130, 204 (2015)

\bibitem {df7}C.R Fadragas and G. Leon, Class. Quantum Grav. 31, 195011 (2014)

\bibitem {df8}T. Nakamura, T. Ikeda, R. Saito, N. Tanahashi and C.-M. Yoo,
Phys. Rev. D 103, 024009 (2021)

\bibitem {df9}S.Kr. Biswas, S. Chakraborty, J.\ Dutta and S. Chakraborty,
Phys. Rev. D 95, 103009 (2017)

\bibitem {df10}M. Kerachian, G. Acquaviva and G. Lukes-Gerakopoulos, Phys.
Rev. D 101, 043535 (2020)

\bibitem {Zhao}D. Zhao, Covariant formulation of f(Q) theory, Eur. Phys. J. C
82, 303 (2022)

\bibitem {Hohmann}M. Hohmann, General covariant symmetric teleparallel
cosmology, Phys. Rev. D 104 124077 (2021)

\bibitem {Heis2}F. D' Ambrosio, L. Heisenberg and S. Kuhn, Revisiting
cosmologies in teleparallelism, Class. Quantum Grav. 39 025013 (2022)

\bibitem {dyn1}W. Khyllep, A. Paliathanasis and J. Dutta, Phys. Rev. D\ 103,
103521 (2021)

\bibitem {dyn2}C. Bohmer, E. Jensko and R. Lazkoz, Universe 9, 166 (2023)

\bibitem {dyn3}H. Shabani, A.\ De and T.-H. Loo, Phase-space analysis of a
novel cosmological model in f(Q) theory, (2023) [arXiv:2304.02949]

\bibitem {self}N. Dimakis, M. Roumeliotis, A. Paliathanasis, P.S.
Apostolopoulos and T. Christodoulakis, Phys.\ Rev. D 106, 123516 (2022)
\end{thebibliography}
\end{document}